\ifx\pdfoptionpdfminorversion\undefined
  \newcount\pdfoptionpdfminorversion
\fi

\documentclass{nato}  
\usepackage{txfonts}
\usepackage{sprrefn} 
\usepackage[draft]{hyperref}
\newdisplay{guess}{Conjecture}
\begin{document}
\firstpage{1}

\begin{opening}
\title{A two-fractal overlap model of earthquakes}
\author{Bikas K Chakrabarti and Arnab Chatterjee\\
\email{bikask.chakrabarti@saha.ac.in,arnab.chatterjee@saha.ac.in}}
\runningauthor{Bikas K Chakrabarti and Arnab Chatterjee}
\runningtitle{Two-fractal overlap models of earthquakes}
\institute{Theoretical Condensed Matter Physics Division and\\
Centre for Applied Mathematics and Computational Science,\\
Saha Institute of Nuclear Physics, 1/AF Bidhannagar, Kolkata 700064, India}

\begin{abstract}
We introduce here the two-fractal model of earthquake dynamics. As the 
fractured surfaces have self-affine properties, we consider the solid-solid
interface of the earth's crust and the tectonic plate below as fractal 
surfaces. The overlap or contact area between the two surfaces give a measure
of the stored elastic energy released during a slip. The overlap between
two fractals change with time as one moves over the other and we show that
the time average of the overlap distribution follows a Gutenberg-Richter
like power-law, with similar exponent value.
\end{abstract}
\keywords{earthquake, fractals, Cantor sets, Gutenberg-Richter law}

\end{opening}

\section{Introduction}
\noindent 
The earth's solid outer crust, about 20 kilometers in
average thickness, rests on the tectonic shells. 
Due to the high temperature-pressure phase changes and the consequent 
powerful convective flow in the earth's mantle (a fluid of very high density), 
at several hundreds of kilometers of depth, the tectonic shell, divided into 
a small number (about ten) of  mobile plates, has relative velocities of the 
order of a few centimeters per year 
\cite{Gutenberg:1954,Kostrov:1989,Scholz:1990}. 
Over several tens of years,  
enormous elastic strains develop on the earth's
crust when sticking (due to the solid-solid friction) to the
moving tectonic plate. When sudden slips occur between the crust and
the tectonic plate, these stored elastic energies are
released in `bursts', causing the damages during the earthquakes.

Earthquakes occur due to fault dynamics in the lithosphere. A
geological fault is created by a fracture in the rock layers, and is 
comprised of the rock surfaces in contact. 
The two parts of the fault are in very slow relative
motion which causes the surfaces to slide. 
Because of the uniform motion of the tectonic plates, the
elastic strain energy stored in a portion of the crust (block),
moving with the plate relative to a `stationary' neighboring part
of the crust, can vary only due to the random strength of
the solid-solid friction between the crust and the plate. 
A slip occurs when the accumulated stress exceeds the resistance 
due to the frictional force.
The potential energy of the strain is thereby released,
causing an earthquake.
As mentioned  before, the observed distribution of the elastic
energy release in various earthquakes seems to follow a power law.

The slip is eventually stopped by friction and stress starts developing again. 
Strain continues to develop till the fault surfaces again slip. 
This intermittent stick-slip process is the essential characteristic feature 
of fault dynamics. The overall distribution of earthquakes, including 
main shocks, foreshocks and aftershocks, is given by the Gutenberg-Richter
law \cite{Gutenberg:1954,Gutenberg:1944}:
\begin{equation}
\log_{10} \mathrm{Nr} ({\cal M}>M) = a - b \: M
\label{eq:Gutenberg-Richter-law}
\end{equation}
where, $\mathrm{Nr}(\mathcal{M}>M)$ denotes the number (or, the
frequency) of earthquakes of magnitudes ${\cal M}$ that are greater than a
certain value $M$. The constant $a$ represents the total number of
earthquakes of all magnitudes: $a = \log_{10} \mathrm{Nr} (\mathcal{M}>0)$
and the value of the coefficient $b$ is presumed to be universal.
In an alternative form, the Gutenberg-Richter law is expressed as a relation
for the number (or, the frequency) of earthquakes in which the energy
released $\mathcal{E}$ is greater than a certain value $E$:
\begin{equation}
\mathrm{Nr} ({\cal E}>E) \sim  E^{-b/\beta},\;\; \mathrm{suggesting}\;\;
\mathrm{Nr} (E) \sim E^{-\gamma},
\label{eq:G-R-power-law}
\end{equation}
for the number density of earthquakes, where $\gamma= 1+ b/\beta$. 
The value of the exponent $\gamma$ is generally observed to be around unity
\cite{Knopoff:2000}; see also {\tt http://web.cz3.nus.edu.sg/\verb+~+chenk/gem2503\_3/notes7\_1.htm}.

\indent One class of models for simulating earthquakes is based on the
collective motion of an assembly of connected elements that are driven slowly,
of which the block-spring model due to Burridge and
Knopoff \cite{Burridge:1967} is the prototype. The Burridge-Knopoff model and
its variants \cite{Carlson:1994,Olami:1992} have the stick-slip dynamics
necessary to produce earthquakes. The underlying principle in this class of
models is self-organized criticality \cite{Bak:1997}.

Another class of models for simulating earthquakes is
based on overlapping fractals, which will be discusses in details in the next
sections. 

\section{Fractals}
\begin{figure}
\centerline{\includegraphics[width=3.2in]{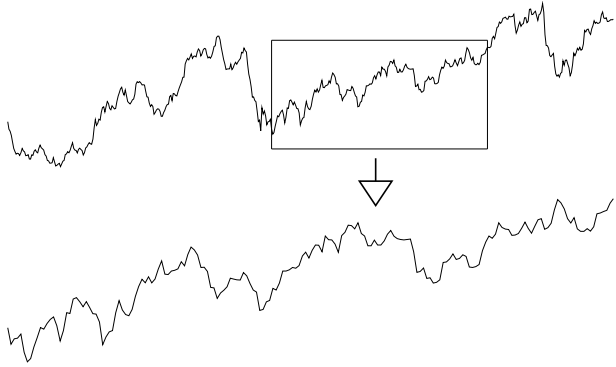}}
\caption[]{Fractal structure of Black sea coastline at I\c{s}ik (Istanbul), 
represented actually by a time series of stock price. This shows the remarkable
self-similarity (examplified by the blow up of the segment in the box)
involved in many such natural processes.}%
\label{deadsea}
\end{figure}

\begin{figure}
\centerline{\rotatebox{270}{\includegraphics[width=2.2in]{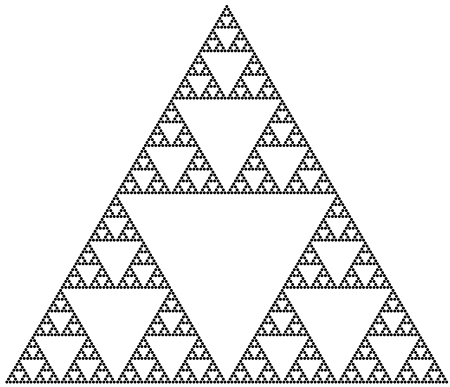}}}
\caption[]{Sierpinski gasket (at generation number $n=5$), having fractal 
dimension $\log 3/\log 2$ (when $n \to \infty$).}%
\label{gasket}
\end{figure}

A fractal is a geometrical object that displays self-similarity on all scales. 
For random fractals, the object need not exhibit exactly the same structure 
at all scales, but the same `type' of structures appear on all scales.
For example, the Black sea coastline measured with different length rulers
will show differences: the shorter the ruler, the longer the `length' measured
but not exactly following the ratio of the inverse of the lengths of the 
rulers; greater than that! This is because more structures come into play
at lower length scales; or in other words, the coastline is not really a `line' 
but rather a fractal having dimension greater than unity. Looking at smaller 
and smaller length-scales, one can find self-similar structures (see Fig.
\ref{deadsea}).

\begin{figure}
\centerline{\rotatebox{0}{\includegraphics[width=3.0in]{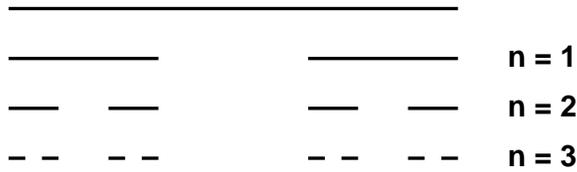}}}
\caption[]{Construction process of Cantor set is shown (upto $n=3$). 
The set becomes a fractal when $n \to \infty$.}
\label{cantor}
\end{figure}

One can easily construct such regular fractals as carpets or gaskets (see 
Fig. \ref{gasket}). In Fig \ref{gasket}, a basic unit of equilateral triangle
with an inner triangle obtained by joining the mid-points of each side
and keeping the inner space void, one constructs a gasket. At each step,
as the length of each side changes by a factor $L=2$, the mass of
the fractal changes by a factor $M (=L^{d_f}) = 3$, giving therefore
the fractal dimension of the object to be $d_f = \log 3/\log 2$. This 
object in Fig. \ref{gasket} of course represents a non-random fractal. However,
a random fractal (as in Fig. \ref{deadsea}) can be easily constructed if 
the void is not always at the center but at any of the $4$ triangles at 
random for each generation; the (mass) dimension $d_f$ remain the same.
A similar fractal of dimension $\log 2/\log 3$ can be constructed
as the Cantor set shown in Fig. \ref{cantor}.
Starting from a set of all real numbers from $0$ to $1$ one removes the 
subset from the middle third, so that for each $l=3$, the mass (size)
of the set $m=2$ and hence the dimension as $n \to \infty$.
This void subset can again be randomly chosen, giving a random Cantor set,
with same $d_f$.

\section{Fractal overlap model of earthquake}
Overlapping fractals form a whole class of models to simulate
earthquake dynamics. These models are motivated by the
observation that a fault surface, like a fractured surface 
\cite{Chakrabarti:1997}, is a fractal object
\cite{Okubo:1987,Scholz:1989,Sahimi:1993}.
Consequently a fault may be viewed as a pair of overlapping fractals.
Fractional Brownian profiles have been commonly used as models of fault
surfaces \cite{Brown:1985,Sahimi:1993}.
In that case the dynamics of a fault is represented by one Brownian
profile drifting on another and each intersection of the two profiles
corresponds to an earthquake \cite{Rubeis:1996}. However
the simplest possible model of a fault $-$ from the fractal point of
view $-$ was proposed by Chakrabarti and Stinchcombe \cite{Chakrabarti:1999}.
This model is a schematic representation of a fault by a pair of dynamically
overlapping Cantor sets. It is not realistic but, as a system of overlapping
fractals, it has the essential feature. Since the Cantor set is a fractal
with a simple construction procedure, it allows us to study in detail the
statistics of the overlap of one fractal object on another.
The two fractal overlap magnitude changes in time as one fractal moves
over the other. The overlap (magnitude)
time series can therefore be studied as a model time series of earthquake
avalanche dynamics \cite{Carlson:1994}.

The statistics of overlaps between two fractals is
not studied much yet, though their knowledge is often required in various
physical contexts. It has been established recently that since
the fractured surfaces have got well-characterized self-affine properties, 
the distribution of the elastic energies
released during the slips between two fractal surfaces (earthquake
events) may follow the overlap distribution of two self-similar fractal
surfaces \cite{Chakrabarti:1999,Pradhan:2003}. 
Chakrabarti and Stinchcombe \cite{Chakrabarti:1999} 
had shown analytically by renormalization group calculations that for 
regular fractal overlap (Cantor sets and carpets) the contact area 
distribution $\rho(s)$ follows a simple power law decay:
\begin{equation}
\rho (s) \sim s^{- \tilde{\gamma}}; \; \tilde{\gamma} = 1.
\label{eq:GR}
\end{equation}
Study of the time ($t$) variation of contact area (overlap)
$s(t)$ between two well-characterized fractals having the same
fractal dimension as one fractal moves over the other with constant
velocity, has revealed some features which can be
utilized to predict the `large events' \cite{Pradhan-K:2004}.

\begin{figure}
\centerline{\rotatebox{0}{\includegraphics[width=4.5in]{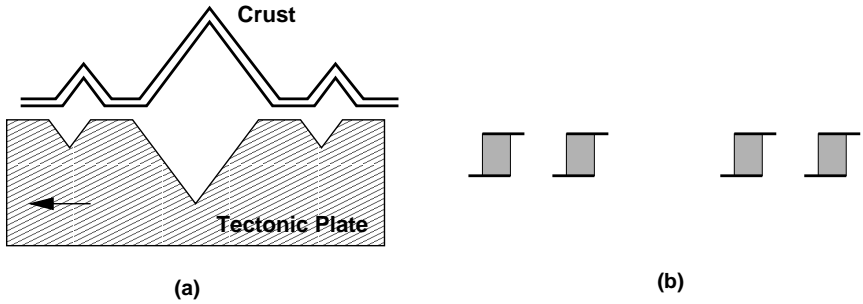}}}
\caption[]{(a) Schematic representations of a portion of the
rough surfaces of the earth's crust and the supporting (moving)
tectonic plate. (b) The one dimensional projection of the surfaces
form Cantor sets of varying contacts or overlaps ($s$) as one surface
slides over the other.}
\label{crust}
\end{figure}

\begin{figure}
\centerline{\rotatebox{0}{\includegraphics[width=3.8in]{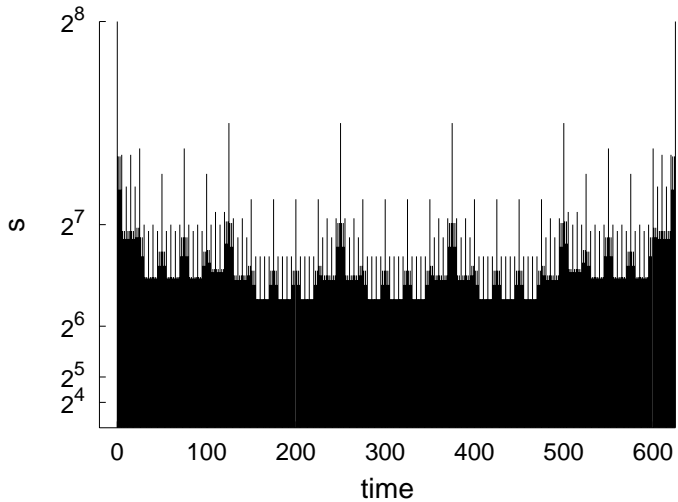}}}
\caption[]{For two Cantor sets (as shown in Fig. \ref{crust}(b); $n=2$), 
one moving uniformly over the other (periodic boundary conditions), 
the total measure of the shaded region 
contribute to the overlap $s$; the time variation of which is shown (for $n=4$).}
\label{tseries}
\end{figure}

Bhattacharyya \cite{Bhattacharyya:2005} has recently studied this overlap
distribution for two Cantor sets with periodic boundary conditions and 
each having dimension $\log 2/\log 3$.
It was shown, using exact counting, that if $s \equiv 2^{n-k}$ ($n$ is the
generation number) then the probability $\tilde{\rho}(s)$ to get an overlap $s$
is given by a binomial distribution \cite{Bhattacharyya:2005}
\begin{equation}
\label{binomial}
\tilde{\rho}(2^{n-k}) = \left ( \begin{array}{c} n\\ n-k \end{array} \right )
 \left ( {1 \over 3} \right )^{n-k} \left ( {2 \over 3} \right )^k
\; \sim \exp(-r^2/n); \; r \to 0,
\end{equation}
where $r^2 = \left[ \frac{3}{2} \left( \frac{2}{3} n - k\right)\right]^2$.
Expressing therefore $r$ by $\log s$ near the maxima of $\tilde{\rho}(s)$, one
can again rewrite (\ref{binomial}) as
\begin{equation}
\label{dtildes}
\tilde{\tilde{\rho}}(s) \sim \exp \left( - \frac{(\log s)^2}{n} \right); \; n \to \infty.
\end{equation}
Noting that $\tilde{\rho}(s) d(\log s) \sim \rho(s) ds$, we find
$\rho(s) \sim s^{-\tilde{\gamma}}$, $\tilde{\gamma}=1$, as in (\ref{eq:GR})
as the binomial or Gaussian part becomes a very weak function of $s$
as $n \to \infty$ \cite{Bhattacharyya:2005a}. It may be noted that 
this exponent value $\tilde{\gamma}=1$ is independent of the
dimension of the Cantor sets considered (here $\log 2/\log 3$) or for that
matter, independent of the fractals employed.
It also denotes the general validity of (\ref{eq:GR}) even for disordered
fractals, as observed numerically \cite{Pradhan:2003,Pradhan-K:2004}.

Identifying the contact area or overlap $s$ between the self-similar
(fractal) crust and tectonic plate surfaces as the stored elastic energy $E$
released during the slip, the distribution (\ref{eq:GR}), of which a
derivation is partly indicated here,
reduces to the Gutenberg-Richter law (\ref{eq:G-R-power-law}) observed.

\section{Summary}
We introduce here the two-fractal overlap model of earthquake where the
average distribution of the overlaps between the surfaces, as one fractal
(here, Cantor set) moves over the other gives the Gutenberg-Richter like
distribution (\ref{eq:GR}), with $\tilde{\gamma}=1$ exactly in the model.
We note that this is an exactly solvable model of earthquake dynamics
and the result for the distributions compare favorably with
the observations.

\acknowledgements
We are grateful to P. Bhattacharyya, P. Chaudhuri, M. K. Dey, S. Pradhan, 
P. Ray and R. B. Stinchcombe for collaborations in various stages of
developing this model.



\theendnotes

\end{document}